\documentclass[twocolumn,showpacs,preprintnumbers,amsmath,amssymb,prl]{revtex4}

\usepackage{graphicx}
\usepackage{dcolumn}
\usepackage{bm}

\begin{document}
\preprint{UNLV-HiPSEC-M04-01}
\title{Condensate-profile asymmetry of a boson mixture in a disk-shaped
harmonic trap}
\author{Hong Ma and Tao Pang}
\affiliation{Department of Physics, University of Nevada, Las Vegas, Nevada
89154-4002}
%
%
\begin{abstract}
A mixture of two types of hard-sphere bosons in a disk-shaped harmonic trap is
studied through path-integral quantum Monte Carlo simulation at low
temperature.  We find that the system can undergo a phase transition to break
the spatial symmetry of the model Hamiltonian when some of the model parameters
are varied.  The nature of such a phase transition is analyzed through the
particle distributions and angular correlation functions.  Comparisons are made
between our calculations and the available mean-field results on similar
models.  Possible future experiments are suggested to verify our findings.
\end{abstract}
\maketitle
 
Mixtures of different cold atoms are studied intensively by both theorists and
experimentalists after the first experimental observation of a stable double
condensate formed from two different spin states of $^{87}$Rb~\cite{my97}.
Interesting phenomena, such as the formation of molecules, can be directly
realized in these mixtures~\cite{he03,re03}.  Furthermore these mixtures form a
rich set of systems that can test many-body theory and novel ideas at extremely
low temperature and quantum limits~\cite{bu03,du03,gu03}.

One of the intriguing features predicted to occur in such systems, based on the
solutions of the Gross--Pitaevskii
equation~\cite{es99,sv03,ao98,oh98,ch99,oh99,ki02,sh04}, is the broken spatial
symmetry of the Hamiltonian in the condensate profiles of individual species in
the system.  This is purely a quantum phenomenon because it is not expected in
the particle distributions of the corresponding classical systems.  It is known
that a double condensate can form an inner core from one species and an outer
shell from the other in a spherically symmetric trap~\cite{ho96,ri02}, which is
a result of the competition between correlation and coherence presented in the
system~\cite{ma03}, while preserving the symmetry of the Hamiltonian.  But
this structure may become unstable for certain interaction strength and total
number of particles and the system can undergo a quantum phase transition to a
state with a broken spatial symmetry~\cite{es99,sv03}---a double-condensate
mixture in an anisotropic trap can develop an asymmetry in condensate profiles
along the direction of the weakest confinement, according to the mean-field
studies~\cite{ao98,oh98,ch99,oh99,ki02,sh04}.  But it is not entirely
certain whether these predications are genuine results of the model system or
some artifacts of the mean-field approximation.

In this Letter we report key results from our recent study of the two-component
hard-sphere boson mixtures in a disk-shaped harmonic trap through path-integral
quantum Monte Carlo simulation, which is exact within a controllable sampling
variance.  This method allows us to study many-body quantum systems at finite
temperature accurately and to make precise predications about the model systems
studied.  The technique has been applied successfully to boson cluster in
various traps~\cite{pe98} and to the two-component system in a harmonic trap
that is spherically symmetric~\cite{ma03}.  For a double condensate in a
disk-shaped harmonic trap, we find that each component can undergo a phase
transition and break the spatial symmetry of the Hamiltonian (or that of the
trap potential) when certain parameters of the system, such as the interaction
range, or the aspect ratio of the axial confinement along the $z$ direction and
the radial confinement in the $xy$ plane, are varied.  The results are in
qualitative agreement with the mean-field
studies~\cite{ao98,oh98,ch99,oh99,ki02,sh04} of this model.  In order to
understand this symmetry-breaking phenomenon in detail, we analyze the
influences of the interaction range, temperature, and number of particles of
each species, external potential and elucidate the nature of this phase
transition.

Here we consider a two-component mixture of hard-sphere bosons described by
the model Hamiltonian
\begin{equation}
\mathcal{H}=\mathcal{H}_1+\mathcal{H}_2+\sum_{j>k=1}^N V_{jk},
\end{equation}
where
\begin{equation}
\mathcal{H}_i=-\frac{\hbar^2}{2m_i}\sum_{l=1}^{N_i}\nabla_l^2
+\sum_{l=1}^{N_i} U_i(\mathbf{r}_l)
\end{equation}
is the Hamiltonian of the corresponding noninteracting system under the
trapping potential $U_i(\mathbf{r}) =m_i\omega_\perp^2 (x^2+y^2+\lambda^2
z^2)/2$.  The parameter $\lambda=\omega_z/\omega_\perp$ is a measure of the
aspect ratio between the axial confinement along the $z$
direction and the radial confinement in the $xy$ plane, with
$\omega_\perp=\omega_x=\omega_y$.  The system is in the shape of a flat disk
if $\lambda\gg 1$ or a long cigar if $\lambda\ll 1$, including the limits
$\lambda=1$ (spherically symmetric), $\lambda=0$ (confined along the $z$
axis), and $\lambda=\infty$ (confined in the $xy$ plane).
The interaction between any two particles $V_{jk}$ is taken to be a
hard-sphere potential, modeled after the $s$-wave scattering lengths of the
experimental systems, with $a_{11}$ and $a_{22}$ being the diameters of
the particles in species 1 and 2, respectively, and $a_{12}=(a_{11}+a_{22})/2$
being the potential range between two different particles.
Masses $m_1$ and $m_2$ and total numbers of particles $N_1$ and $N_2$ are
used for each of the two species, respectively, with $N=N_1+N_2$ being the
total number of particles in the system.

We performed simulations by carefully choosing the total number of time
slices and the size of the Monte Carlo steps to ensure a good convergence
of the physical quantities calculated, including the density
$\rho_i(\mathbf{r})$ and the in-plane angular pair-correlation function
$\Gamma_i(\varphi) = \langle\delta(\varphi-|\varphi_l-\varphi_k|)\rangle$
for $l\not=k$. 
In order to examine the condensate profiles against the planar
symmetry of the trap, we set the temperature much lower than the critical
temperature and projected all the particles into the $xy$ plane when we take
snapshots or calculate the average radial density $\rho_i(\rho)$ and angular
correlation function $\Gamma_i(\varphi)$, where $\rho=\sqrt{x^2+y^2}$ is the
in-plane radius and $\varphi$ is the relative azimuthal angle.
The density is
normalized by $2\pi\int_0^\infty\rho_i(\rho)\rho\,d\rho=N_i$.

\begin{figure}[!ht]
\resizebox{2.6in}{!}{\includegraphics{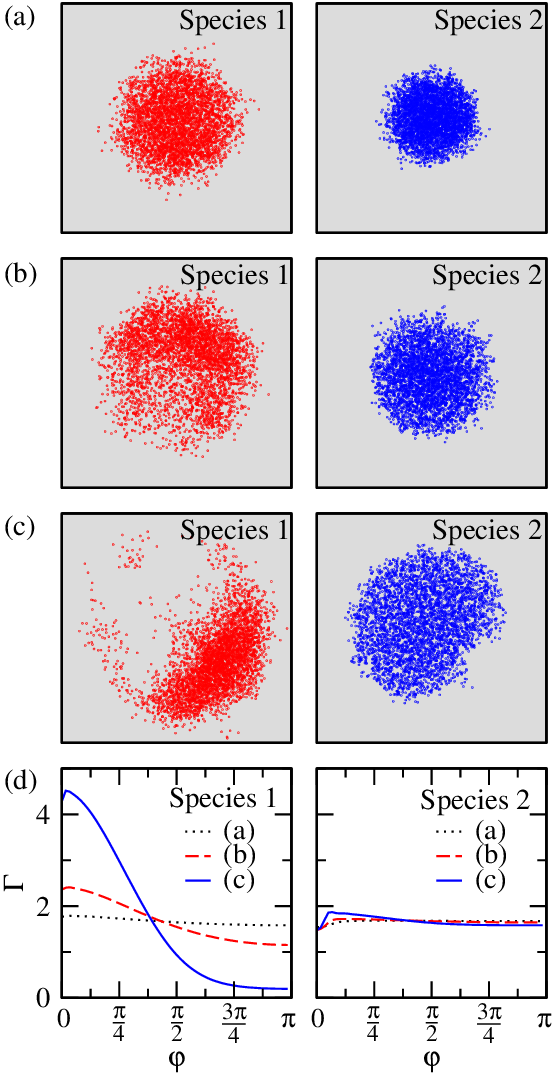}}
\caption{Snapshots of multi-exposures of the particles viewed along the $z$ 
direction for each species in a double-condensate mixture, confined in a trap 
with $m_{2}/m_{1}=4$,
$N_1=N_2=100$, $a_{11}=0.2$, and $a_{22}=0.4$, for different aspect ratio
$\lambda=\omega_z/\omega_\perp$: (a) $\lambda=1$; (b) $\lambda=4$; and (c)
$\lambda=16$.  The corresponding in-plane angular pair-correlation functions
$\Gamma_i(\varphi)$ are shown in (d). The planar symmetry in each of the
two condensates is broken if $\lambda\not=1$; the asymmetry becomes more 
severe when $\lambda$ is further away from 1, especially for the lighter 
species.}
\label{fig1}
\end{figure}

In Fig.~\ref{fig1} we show three sets of simulation results for different
aspect ratio $\lambda$ under the same trapping frequency for both species at
$T/N^{1/3}=0.1$. It is clear that the symmetry
of the Hamiltonian is preserved in the condensate profiles when the trap is
spherically symmetric, that is, $\lambda=1$.  For both $\lambda=4$ and
$\lambda =16$, the profiles are obviously asymmetric with a random orientation
due to the spontaneous symmetry-breaking. The asymmetry becomes more severe 
and the correlation at small angle, as shown in Fig.~\ref{fig1}(d), becomes 
much stronger as the ratio $\lambda$ increases.

The symmetry-breaking in one species compensates the other because
the center of mass of the system, under the given conditions, remains at the 
center of the trap, resulting in a stronger deformation in the lighter 
species.  We performed one more simulation with $\lambda<1$ and did not see 
any symmetry-breaking of the profiles in the $xy$ plane.  This is consistent 
with the findings of previous mean-field calculations~\cite{es99, ki02} based 
on the Gross--Pitaevskii equation, which concluded that the symmetry is broken 
in the direction of the weakest trapping frequency.

\begin{figure}[!ht]
\resizebox{2.6in}{!}{\includegraphics{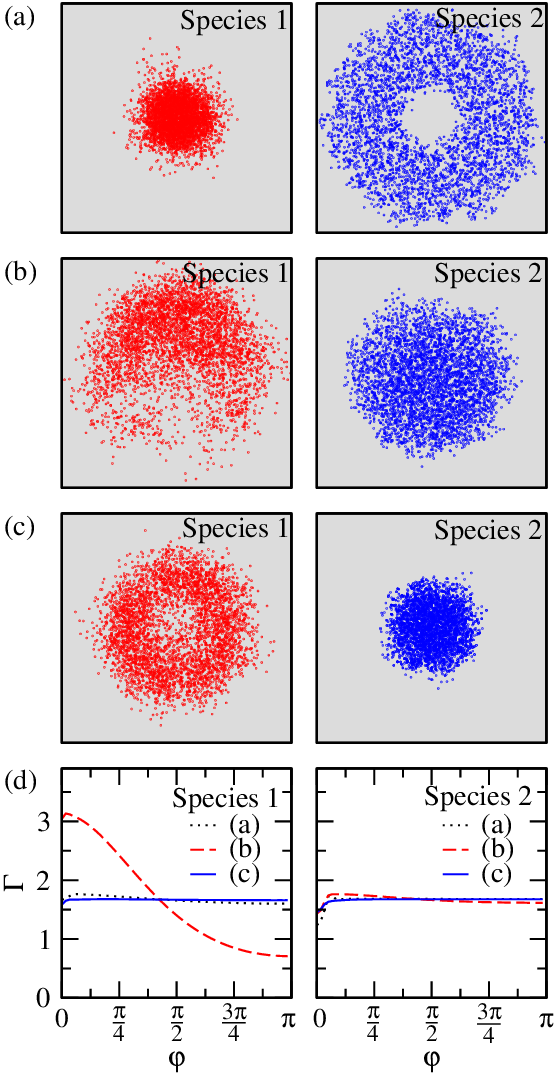}}
\caption{Snapshots of multi-exposures of the particles viewed along the $z$ 
direction for each
species in a double-condensate mixture, confined in a trap with $m_{2}/m_{1}
=4$, $N_{1}=N_{2}=100$, $\lambda=\omega_z/\omega_\perp=16$, and $a_{11}=0.15$
fixed, for different ratio $\eta=a_{11}/a_{22}$: (a) $\eta=1/3$; (b) $\eta
=1/2$; and (c) $\eta=1$. The corresponding in-plane angular pair-correlation
functions $\Gamma_i(\varphi)$ are shown in (d). The planar symmetry of
the particle distributions is broken first and then restored when the ratio is
varied monotonically.}
\label{fig2}
\end{figure}

Let us now examine how the system responses to the change of the ratio
$\eta=a_{11}/a_{22}$, which is a measure of the imbalance of the particle
sizes between the two species.  In Fig.~\ref{fig2} we show three sets of
snapshots with different $\eta$ while having other parameters
$m_2/m_1=4$, $N_1=N_{2}=100$, $\lambda=16$, and $a_{11}=0.15$
fixed.

For the case of $\eta=1/3$, the system shows a condensed core of species
1, surrounded by an outer ring of species 2.  Both condensates appear to have
the planar symmetry of the trap.  When $\eta$ is increased to 1/2, the system
undergoes a quantum phase transition from two separate condensates to a binary
mixture with each breaking the planar symmetry.  This is evident from both
snapshots and the correlation functions shown.  The correlation at small angle 
increases when the symmetry is broken. Note that the condensate of species 1 
is expected more off-centered ( 4 times) because the center of mass
of the whole system must remain at the trap center.  One can see from the
snapshots that the binary phase is formed from the outward movement of species
1 plus the inward movement of species 2.  Therefore, increasing $\eta$ further
can result in exchanging the roles of the two species.  This is precisely
what happens when $\eta$ is increased to 1---the system shows a condensed core
of species 2, surrounded by an outer ring of species 1. The correlation 
at small angle decreases and the planar symmetry of the Hamiltonian gets 
restored in the condensate profiles in the same time. 

\begin{figure}
\resizebox{2.6in}{!}{\includegraphics{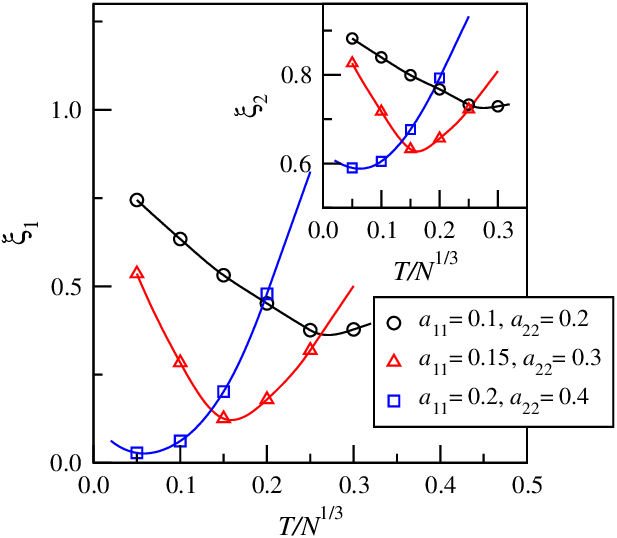}}
\caption{Temperature dependence of the condensate profile asymmetries, measured
from the ratio $\xi=N_{\mathrm{L}}/N_{\mathrm{H}}$, where $N_{\mathrm{L}}$ is
the number of the particles in the low-density half and $N_{\mathrm{H}}$ is the
number of particles in the high-density half, with $\eta=a_{11}/a_{22}=1/2$,
$m_2/m_1=4$, $N_1=N_2=100$, and $\lambda=16$ fixed.  The most severe asymmetry
for each case happens at a temperature that decreases with the interaction
(hard-sphere diameter).}
\label{fig3}
\end{figure}

To investigate the temperature dependence of asymmetry of each condensate, 
we divide the trap into two
equal halves by a plane through the $z$ axis, with the low-density side
containing a minimum number of particles ($N_{\mathrm{L}})$ and the
high-density side containing a maximum number of particles ($N_{\mathrm{H}}$).
Then we examine the ratio $\xi=N_{\mathrm{L}}/N_{\mathrm{H}}$, which is shown
in Fig.~\ref{fig3} for a trap with $\eta=1/2$, $m_2/m_1=4$, $N_1=N_2=100$, and
$\lambda=16$ fixed.

A general feature emerges from all these systems: The asymmetry is depleted
in both the extremely low temperature region and high temperature region.
The causes are however different.  At the extremely low temperature, the
strong coherence in each of the condensates forces the particles to go to
the lowest single-particle states, and hence to form a nearly symmetric
distribution.  This is extremely evident from the systems with a weaker
interaction, for example, in cases with $a_{11}=0.1$ and $a_{11}= 0.15$, where
the symmetry is nearly completely restored when temperature is lower than
$0.05 N^{1/3}$.  At the high temperature, particles gain more thermal
(kinetic) energy and can therefore spread out more in order to distribute
more symmetrically.  More detailed analysis of the data also show that for
weaker interaction cases, the particles from one condensate penetrate into the
other condensate when temperature is increased, whereas for the stronger
interaction cases, the particles from one condensate go around to form an
outer ring. 

The optimal temperature, at which the system shows a minimum 
$N_{\mathrm{L}}/N_{\mathrm{H}}$ (maximum asymmetry) in Fig.~\ref{fig3}, 
decreases with the interaction range.  This is consistent with
our earlier findings~\cite{ma03} that the many-body correlation effect,
originated from the strong interaction, is in competition with the coherence
and it is more difficult to restore the symmetry in the systems with a 
strong interaction.  Another distinctive feature obtained here is that the 
minimum ratio of $N_{\mathrm{L}}/N_{\mathrm{H}}$ is smaller if the interaction 
range is greater.  In other
words, the system with a stronger interaction can reach a higher asymmetry.
  
\begin{figure}
\resizebox{3.4in}{!}{\includegraphics{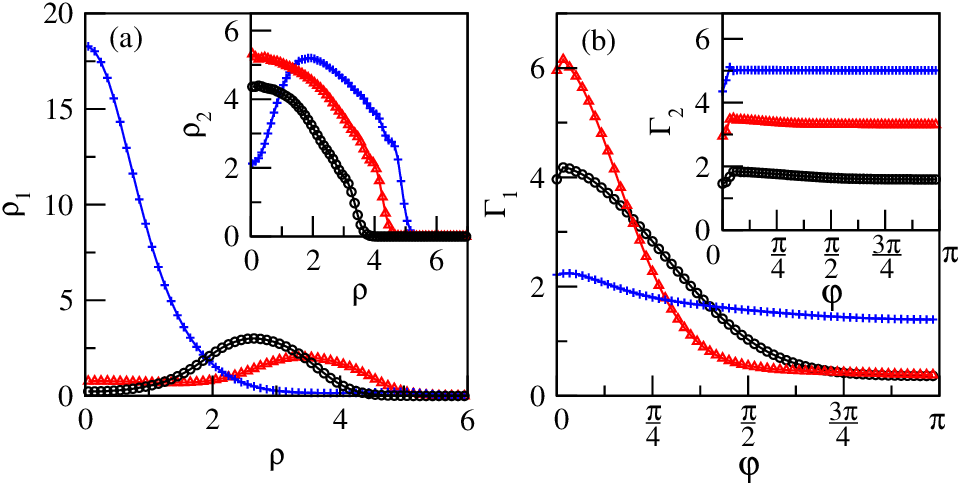}}
\caption{(a) Density profiles $\rho_i(\rho)$ and (b) angular pair-correlation
functions $\Gamma_i(\varphi)$ of the particles confined in a trap with 
$N_1=100$, $m_{2}/m_{1}=4$, $a_{11}=0.2$, $a_{22}=0.4$, and $\lambda=16$ fixed,
for $N_2=100$ (circles), $200$ (triangles), and $300$ (crosses).  As $N_2$
increases, the lighter particles, while maintaining certain asymmetry, contract
and are eventually surrounded by the heavier particles that always appear to be
distributed symmetrically and are pushed out of the center.}
\label{fig4}
\end{figure}

In addition to the interaction and temperature, the structure of a
double-condensate mixture also depends on the total number of particles in
the system.  In Fig.~\ref{fig4}, we show the results of three different $N_2$
while keeping $N_1=100$, $m_2/m_1=4$, $a_{11}=0.2$, $a_{22}=0.4$, and
$\lambda=16$ fixed.

From the correlation functions shown for $N_2=N_1=100$, the particles from
species 1 occupy one side of the trap with a significant asymmetry while the
particles from species 2 stay at the center of trap with a smaller asymmetry,
also seen in Fig.~\ref{fig1} (c).  Quantitatively, one expects that the center
of mass of species 1 is about four times away from the trap center of that of
species 2 in this case because the center of mass of the system is at 
the center of
the trap.  When $N_2$ is increased to 200, the particles in species 2
squeeze each other outward but push some of the particles in species 1 into
the central area and others further off the center.  Finally when $N_2$ is
increased to $300$, particles in species 1 all move to the central area;
both condensates recover the planar symmetry of the trap with particles
in species 2 forming an outer ring, surrounding completely the particles from 
species 1.  However when $N_1$ is increased from 100 to 300 with
$N_2$ kept to be 100, the asymmetries return with each condensate occupying
one side of the trap.

So far we have shown the results with both the species under the same trapping
frequency, which means a different external potential for each species
if $m_2/m_1\not=1$.  In order to examine the influence of external potential,
we can either choose $m_{2}/m_{1}=1$ with the same trapping frequency or
set $m_1/m_2=\omega^2_2/\omega^2_1$.  Such simulations performed for those
systems depicted in Fig.~\ref{fig1} show no broken symmetry as the aspect ratio
$\lambda$ is increased from 1 to 16, although phase separation still occurs 
along
$\rho$ direction.  This is similar to the case of our previous work on the
double-condensate mixture in a spherically symmetric trap~\cite{ma03};  we
also find that the particles with a larger interaction range are in the outer
ring while the other condensate stays at the central area.

The external potential plays an important role in the development of asymmetry
in the systems studied.  For a system with $m_2/m_1=4$ under the same trapping
frequency, the condensates can undergo a symmetry-to-asymmetry transition and
then an asymmetry-to-symmetry transition, as shown in Fig.~\ref{fig2}.
However, for a system with $m_2/m_1=1$, still under the same trapping frequency,
no significant asymmetry can be developed. The system with $a_{11}/a_{22}=0.5$
can have a stable, phase-separated configuration similar to that of
Fig.~\ref{fig2}(a), which is also a stable configuration obtained in a
mean-field approach~\cite{sv03} with $a_{12}/(a_{11}a_{22})^{1/2}\approx 1.06$.
Furthermore, when $a_{11}/a_{22}$ is increased to 1, the two separated
condensates penetrate each other and form a binary mixture with a slight
asymmetry---the double condensates are mirror images of each other because of
the interchangeable nature of the two species under the given conditions.

In summary, our simulations show that asymmetry can occur in a
double-condensate mixture confined in a disk-shaped trap.  The asymmetry
has a strong
dependence on the aspect ratio of the axial confinement along the $z$ direction
and the radial confinement in the $xy$ plane, the ratio of the interaction
ranges $a_{11}/a_{22}$, and the temperature.  Furthermore, the total numbers
of particles in the two species and significant difference in the external
potentials can also affect the structures of the condensates.  The novel
behavior of a double-condensate mixture elucidated in this work can be
verified experimentally with the setup that has created the mixtures and it is
extremely interesting to see such phenomena involving quantum phase transition
and reentry of symmetry of a model Hamiltonian realized in a laboratory.

This work is supported in part by DOE under Cooperative Agreement
DE-FC52-01NV14049 and by NSF under Cooperative Agreement ACI-9619020
through the computing resources provided by the National Partnership for
Advanced Computational Infrastructure at the University of Michigan Center
for Advanced Computing.

\end{document}